\documentstyle[12pt]{article}

\newskip\zatskip \zatskip=0pt plus0pt minus0pt
\def\matth{\mathsurround=0pt}

\def\atversim#1#2{\lower0.7ex\vbox{\baselineskip\zatskip\lineskip
  \zatskip
  \lineskiplimit 0pt\ialign{$\matth#1\hfil##\hfil$\crcr#2\crcr\sim
  \crcr}}}

\input epsf 

\begin{document}

\begin{titlepage}

\hspace*{\fill}\parbox[b]{3.4cm}{MSUHEP-61217 \\ December 1996\\ 
hep-ph/9612454\\
\vspace*{2.0cm}}

\vspace*{-0.5cm}

\begin{center}
\large\bf
A Monte Carlo 
Solution to the BFKL Equation
\end{center}
 
\vspace*{0.3cm}
 
\begin{center}
Carl R. Schmidt \\
Department of Physics and Astronomy\\
Michigan State University\\
East Lansing, MI 48824, USA
\end{center}
 
\vspace*{0.3cm}
 
\begin{center}
\bf Abstract
\end{center}
 
\noindent

A simple solution to the BFKL equation is obtained as a series in the
number of real gluons emitted with transverse momentum greater than some
small cutoff $\mu$.  This solution reveals physics inside the BFKL
ladder which is hidden in the standard inclusive solution,
and lends itself to a straightforward Monte Carlo implementation.
With this approach one can explore new useful physical observables, 
which are shown to be independent of the cutoff $\mu$.
In addition, this approach allows the imposition of kinematic constraints
(such as energy conservation) which are important at finite energies.
The distribution
of $\sum E_\perp$ of particles in a central rapidity bin between two
widely-spaced jets is presented as an example.

\end{titlepage}
 
\baselineskip=0.8cm
 
 
The BFKL (Balitsky-Fadin-Kuraev-Lipatov)
equation \cite{BFKLpapers}, 
which systematically  resums powers of $\alpha_s$ times 
large rapidity intervals or logarithms of
Feynman $x$ in perturbative QCD, has recently moved from the purely 
theoretical realm to the phenomenological arena.  The classic prediction of
the BFKL resummation is the rise of $F_2(x)$ at small $x$ in deep inelastic
scattering.    Unfortunately, due to the resiliency of the parton density 
functions and DGLAP \cite{dglap} evolution equations, this observation 
\cite{hone,zeus} is still open to 
interpretation \cite{catani}.  Alternatively, by tagging jets on both ends 
of a large 
rapidity interval, as suggested by Mueller and Navalet \cite{MN}, it is 
possible to unambiguously isolate the effects of 
the BFKL ladder from the parton density functions.
Predictions of this kind include the rise in the cross section as a
function of the rapidity interval, both in hadron-hadron \cite{MN} and 
lepton-hadron colliders \cite{mueller}.  
In particular, recent preliminary results from H1 \cite{hone} on deep 
inelastic scattering
with a tagged jet at fixed $x_j$ show an intriguing rise in the cross section
with decreasing $x_{\rm BJ}$ that is not well-explained either by a 
fixed-order matrix element calculation \cite{mirzep} 
or by parton shower Monte Carlo 
simulations.  The BFKL prediction \cite{bdrg}, however, seems to be
in good agreement.
Other observables related to the kinematics of the tagging jets, such as the 
decorrelation in azimuthal angle \cite{DDS}, have also been considered.

With the advent of BFKL phenomenology, it has become a necessity to 
understand the range of validity and the errors inherent in the BFKL 
approximations.  The calculation of the next-to-leading logarithmic (NLL) 
corrections to the BFKL matrix elements is currently under way \cite{lip}.
Short of these full NLL corrections, it has been shown in \cite{DDS2} 
that some of the largest corrections to the asymptotic theory
are purely kinematic in origin.  
In the standard solution to the BFKL equation the 
transverse momenta of the ladder gluons are integrated from zero to 
infinity.  However, at physical energies the inclusion of kinematic 
constraints on these integrals can significantly modify the
predictions of the theory, even if the difference is formally subleading
in the asymptotic expansion.  In Ref.~\cite{DDS2} an effective rapidity 
interval was defined as an attempt to include 
some of these kinematic effects in the BFKL calculation.

In this paper we present a new solution to the BFKL equation 
in a form naturally suited for physical interpretation and Monte
Carlo implementation \cite{oedipus}.  This approach also offers a natural
way to impose kinematic constraints and to assess the uncertainties due
to the asymptotic nature of the equation.  
The key ingredient to this solution is the introduction
of a lower cutoff $\mu$ on the transverse momentum of the real 
gluons that are produced.  However, the cutoff is introduced in such 
a way that, for sufficiently small $\mu$, any infrared safe
observable is independent of the cutoff.  In the inclusive case
the Monte Carlo method generates an exact answer, identical to the
known solution in the literature, with no arbitrary parameters.
In addition, new experimental observables which depend on the 
observation of the ladder gluons can be obtained \cite{martin}.

Let us begin with a general description of a semi-hard
process, where the parton-parton center-of-momentum energy $\sqrt{\hat
s}$ is much larger than the typical momentum transfer scale $Q$.  In 
this limit the partonic cross section factorizes into the form
\begin{equation}
{d\hat\sigma\over d^2\mbox{\boldmath $p_{a\perp}$}
d^2\mbox{\boldmath $p_{b\perp}$}}\ =\ V_a(p_{a\perp}^2)\,
f(y_{ab},\mbox{\boldmath $p_{a\perp}$},\mbox{\boldmath 
$p_{b\perp}$})\, V_b(p_{b\perp}^2) \label{semihard}\ .
\end{equation}
A physical interpretation of this form of the cross section is represented
in Fig.~1.   The process consists of two distinct scatterings, which 
occur at widely-separated rapidities, $y_a$ and $y_b$,
and small transverse momenta $p_{a\perp}\!=\!
|\mbox{\boldmath $p_{a\perp}$}|
\sim p_{b\perp}\!=\!|\mbox{\boldmath $p_{b\perp}$}|\sim Q$.
Each scattering form factor, $V_a(p_{a\perp})$ and $V_b(p_{b\perp})$, 
depends only on the 
transverse momentum which flows into its particular vertex.
The precise form of the form factors,
however, depends on the specific partons involved in the scatterings.  
The function $f(y_{ab},\mbox{\boldmath $p_{a\perp}$}
,\mbox{\boldmath $p_{b\perp}$})$, which connects the two scatterings, 
is essentially a propagator which allows 
$\mbox{\boldmath $p_{b\perp}$}$ to flow to $\mbox{\boldmath $p_{a\perp}$}$
by emitting gluons over a rapidity interval $y_{ab}=y_a-y_b
\sim \ln(\hat s/p_{a\perp}p_{b\perp})$.
This function is universal, and it is our object of interest.

The form of the cross section given in (\ref{semihard}) is valid in the
limit of large $y_{ab}$.  In that limit the perturbative contributions to
$f(y_{ab},\mbox{\boldmath $p_{a\perp}$}
,\mbox{\boldmath $p_{b\perp}$})$ that are leading in $\alpha_s y_{ab}$
can be resummed systematically with the aid of the BFKL equation.
We now present the equation and describe the physics that it 
encodes.  It can be written
\begin{eqnarray}
\lefteqn{{\partial f(y_{ab},\mbox{\boldmath $p_{a\perp}$},\mbox{\boldmath 
$p_{b\perp}$})\over\partial y_a}\ =\  
{\bar\alpha_s\over\pi}\int{d^2\mbox{\boldmath $k_\perp$}
\over k_\perp^2}\Biggl[
f(y_{ab},
\mbox{\boldmath $p_{a\perp}$}\!+\!\mbox{\boldmath $k_\perp$},
\mbox{\boldmath $p_{b\perp}$})}\label{bfkl}\\
&&\qquad\qquad\qquad\qquad\qquad\qquad\quad
-{p_{a\perp}^2\over k_\perp^2+
(\mbox{\boldmath $p_{a\perp}$}\!+\!\mbox{\boldmath $k_\perp$})^2}
f(y_{ab},\mbox{\boldmath $p_{a\perp}$},\mbox{\boldmath $p_{b\perp}$})
\Biggr]\nonumber \ ,
\end{eqnarray}
where $\bar\alpha_s=\alpha_s N_c/\pi$.
The boundary condition for the equation is 
\begin{equation}
f(0,\mbox{\boldmath $p_{a\perp}$},\mbox{\boldmath 
$p_{b\perp}$})\ =\  
{1\over2}\delta^{(2)}
(\mbox{\boldmath $p_{a\perp}$}+\mbox{\boldmath $p_{b\perp}$})
\ ,
\label{bc}
\end{equation}
which corresponds to no gluon emissions and enforces conservation of 
transverse momentum.  The first term on the
right-hand side of the BFKL equation (\ref{bfkl}), upon iterating from
the boundary condition $n$ times, gives the squared 
amplitude for producing $n$ real gluons, in the approximation that the gluons
are well-separated in rapidity.  The amplitude with three gluon emissions 
is represented by the Feynman diagram
in Fig.~2.  The second term on the right hand side of (\ref{bfkl}) 
gives the virtual 
corrections which ``reggeize'' the $t$-channel gluon propagators, 
represented by the heavy solid lines in the figure.

An important point here is that the singularities in the 
real and virtual terms
of the integrand cancel as $k_\perp$ becomes small.  Thus, we can cut off the 
integral at $k_\perp=\mu$ for both the real and virtual gluons,
and the neglected contribution will vanish as $\mu
\rightarrow 0$.  With this cutoff we can explicitly do the integration
over the virtual gluon corrections, obtaining:
\begin{eqnarray}
{\partial f(y_{ab},\mbox{\boldmath $p_{a\perp}$},\mbox{\boldmath 
$p_{b\perp}$})\over\partial y_a} & =&  
{\bar\alpha_s\over\pi}\int{d^2\mbox{\boldmath $k_\perp$}
\over k_\perp^2} f(y_{ab},
\mbox{\boldmath $p_{a\perp}$}\!+\!\mbox{\boldmath $k_\perp$},
\mbox{\boldmath $p_{b\perp}$})\\
&&+\,\bar\alpha_s \ln(\mu^2/p_{a\perp}^2)
f(y_{ab},\mbox{\boldmath $p_{a\perp}$},\mbox{\boldmath $p_{b\perp}$})
\ +\ {\cal O}(\mu^2/p_{a\perp}^2)\nonumber \ ,
\label{bfklii}
\end{eqnarray}
where the integral over real gluons is restricted to 
$k_\perp>\mu$.
From hereon we neglect the terms of ${\cal O}(\mu^2/p_{a\perp}^2)$.
Then we can simplify this equation further by making the substitution
\begin{equation}
f(y_{ab},\mbox{\boldmath $p_{a\perp}$},\mbox{\boldmath $p_{b\perp}$})
\ =\ 
\Biggl({\mu^2\over p_{a\perp}^2}\Biggr)^{\bar\alpha_s y_{ab}
}\tilde{f}
(y_{ab},\mbox{\boldmath $p_{a\perp}$},\mbox{\boldmath $p_{b\perp}$})\ ,
\label{ftilde}
\end{equation}
which leaves the following equation \cite{KLM}:
\begin{equation}
{\partial \tilde{f}(y_{ab},\mbox{\boldmath $p_{a\perp}$},\mbox{\boldmath 
$p_{b\perp}$})\over\partial y_a}\ =\  
{\bar\alpha_s \over\pi}\int{d^2\mbox{\boldmath $k_\perp$}
\over k_\perp^2}
\Biggl({p_{a\perp}^2\over(\mbox{\boldmath $p_{a\perp}$}
\!+\!\mbox{\boldmath $k_\perp$})^2}\Biggr)^{\bar\alpha_s y_{ab}}
\tilde{f}(y_{ab},
\mbox{\boldmath $p_{a\perp}$}\!+\!\mbox{\boldmath $k_\perp$},
\mbox{\boldmath $p_{b\perp}$})\ ,
\label{bfkliii} 
\end{equation}
with the same boundary condition (\ref{bc}).

In (\ref{ftilde}) we see that the dependence of the virtual corrections
on $\mu$ 
simply exponentiates into an overall factor.  Starting with the boundary
condition, we can iterate equation (\ref{bfkliii})
 to obtain a series solution 
$f=\sum_{n=0}^\infty f^n$, where the $n$th term 
is the contribution from the emission of $n$
real gluons having $k_\perp>\mu$.
This contribution can be written as a product of integrals
over the phase space of each gluon in the ladder:
\begin{eqnarray}
\lefteqn{f^n(y_{ab},\mbox{\boldmath $p_{a\perp}$},
\mbox{\boldmath $p_{b\perp}$})\ =\ \int
\prod_{i=1}^n\Biggl[\bar\alpha_s
dy_i{dk_{i\perp}^2\over k_{i\perp}^2}
{d\phi_i\over2\pi} \Biggl({\mu^2\over q_{i\perp}^2}\Biggr)^
{\bar\alpha_s y_{i+1,i}}\Biggr]
\Biggl({\mu^2\over q_{0\perp}^2}\Biggr)^{\bar\alpha_s y_{1,0}} 
}  \label{soln} \\ 
&&\qquad\qquad\qquad\qquad\qquad\qquad\qquad\qquad\qquad\qquad\qquad
\times\ 
{1\over2}\delta^{(2)}
(\mbox{\boldmath $p_{a\perp}$}+\mbox{\boldmath $q_{n\perp}$})\nonumber
\end{eqnarray}
with
\begin{equation}
\mbox{\boldmath $q_{j\perp}$}\ =\ 
\mbox{\boldmath $p_{b\perp}$} +
\sum_{i=1}^j\mbox{\boldmath $k_{i\perp}$}
\label{qn}
\end{equation}
and $\mbox{\boldmath $q_{0\perp}$}=\mbox{\boldmath $p_{b\perp}$}$.
The integrals in rapidity are ordered with $y_b\equiv y_0<y_1<\cdots
<y_n<y_{n+1}\equiv y_a$.  Written in this manner, the solution
recovers the simple interpretation as the emission of real gluons by
the exchange of ``reggeized'' gluons in the $t$-channel as in
Fig.~2, with $\mu$ the infrared cutoff  to the gluon Regge trajectory.
For clarity, we emphasize that only the gluons exchanged in the $t$-channel 
are ``reggeized'', while the emitted partons are standard gluons.

Each term in the series is positive definite.  Therefore, it
is straightforward to implement this solution as a Monte Carlo 
simulation.  Given $\mbox{\boldmath $p_{b\perp}$}$ and the rapidity
interval $y_a-y_b$, we first sample the distribution in the number $n$ of 
gluons in the ladder. Next, we produce the four-momenta of the $n$ gluons 
successively as given by the distribution (\ref{soln}).  Finally, we fix 
$\mbox{\boldmath $p_{a\perp}$}$ by conservation of transverse momentum.  
In practice the events are produced using approximate distributions and
are then reweighted.

To understand more clearly the cutoff dependence of this solution, let us 
calculate the
quantity $F=\int f\,dp_{a\perp}^2d\phi_a$, where the integral just fixes 
$\mbox{\boldmath $p_{a\perp}$}$ via the $\delta$-function in (\ref{soln}).
For simplicity of the discussion here, it is convenient to consider a 
modified equation,
obtained by replacing $q_{i\perp}\rightarrow p_{b\perp}$ everywhere 
in (\ref{soln}) and by 
setting the upper limit on the integrations to $k_\perp^2=p_{b\perp}^2$.  
Now the nested integrals can be done analytically, the series can be summed,
and the dependence on $\mu$ vanishes identically.
In fact we obtain $F=1$, and the distribution in the number of ladder gluons 
is just a Poisson distribution with mean 
\begin{equation}
\langle n\rangle\ =\ \bar\alpha_s y_{ab}
\ln{(p_{b\perp}^2/\mu^2)}\ .\label{nmean}
\end{equation}
The physical significance of the cutoff is now apparent.  As we lower
 $\mu$, the number of gluons emitted in the rapidity
interval grows logarithmically.  However, the average $k_{\perp}$ of
each gluon is reduced in such a way that, for suitable infrared-finite 
observables, the dependence on the cutoff vanishes.    
In the exact solution (\ref{soln}) the distribution in the number of ladder 
gluons will differ somewhat, but the qualitative features of this discussion 
still apply.  

We now present in Fig.~3 a plot of the Monte Carlo solution compared to the
standard BFKL solution \cite{BFKLpapers}, 
\begin{equation}
f(y_{ab},\mbox{\boldmath $p_{a\perp}$},\mbox{\boldmath 
$p_{b\perp}$})\ =\ {1\over(2\pi)^2 p_{a\perp}p_{b\perp}}
\sum_{n=-\infty}^{\infty} e^{in(\phi_{ab}-\pi)}
\int_{-\infty}^{\infty} d\nu\, e^{\omega(n,\nu)\, y_{ab}}
\,(p_{a\perp}^2/p_{b\perp}^2)^{i\nu}\ ,\label{standard}
\end{equation}
with $\phi_{ab}=\phi_a-\phi_b$ and
\begin{equation}
\omega(n,\nu)\ =\ 2\bar\alpha_s
\bigl[\psi(1) - {\rm Re}\,\psi ({|n|+1\over 2} +i\nu) \bigr]\ ,
\label{eigen}
\end{equation}
where $\psi$ is the logarithmic derivative of the Gamma function.
In this figure we fix $p_{b\perp}=50$ GeV and $y_{ab}=4$, 
and we plot $\int f\,d\phi$ as a function of $p_{a\perp}$.
The Monte Carlo solution agrees with the standard solution (\ref{standard})
for this curve as long as $\mu$ is smaller than the bin size
used near the peak.

Next, we calculate an observable for the Tevatron at $\sqrt{s}=1800$ GeV
that exhibits the full potential of the Monte Carlo approach to BFKL 
physics.  We tag on the two jets $a$ and $b$ with the largest and 
smallest rapidities with $p_{a\perp},p_{b\perp}>20$ GeV.  Then we sum
the $E_{\perp}$ of all the particles within a bin in rapidity and 
plot the distribution in $\Sigma_{\perp}=\sum E_{\perp}$.  In Fig.~4 we
plot the distribution of $\Sigma_{\perp}$ summed in the bin 
$|y|<0.5$ for fixed values of 
$y_{a}=2.5$ and $y_{b}=-2.5$ for the tagging jets.  In this 
calculation we have also made an improvement to the BFKL
prediction by including the kinematic contribution of all of the
physically-produced particles to the Feynman-$x$ parameters
in the parton density functions of the proton and anti-proton.  In 
practice, this makes a large effect due to the constraint of total 
energy conservation imposed by the parton density functions.  More 
details of this kinematic improvement and other phenomenological 
results will be reported in an expanded paper \cite{future}. 

In conclusion, we have presented a solution to the BFKL equation
for large rapidity intervals as a series in the number of real gluons
emitted above a cutoff $\mu$.  We have incorporated this
solution into a Monte Carlo simulation, and we have shown that 
it reproduces exactly the BFKL dynamics with no dependence on the 
cutoff for small $\mu$.  An advantage of a Monte Carlo 
solution to BFKL is that it allows one to study the effects of the 
gluons emitted in the middle of the ladder and even to make 
experimental cuts on these ladder gluons.  Finally, with the Monte 
Carlo simulation it is also possible to improve the convergence of the 
resummation by including kinematical effects exactly in the cross section.

The author would like to thank Mike Albrow, Vittorio Del Duca, 
Terry Heuring, Joey Huston, Harry Weerts, and 
C.-P. Yuan for useful conversations and Wu-Ki Tung for a critical
reading of this manuscript.

\newpage

\begin{figure}
\vskip-5.0cm
\epsfysize=16cm
\centerline{\epsffile{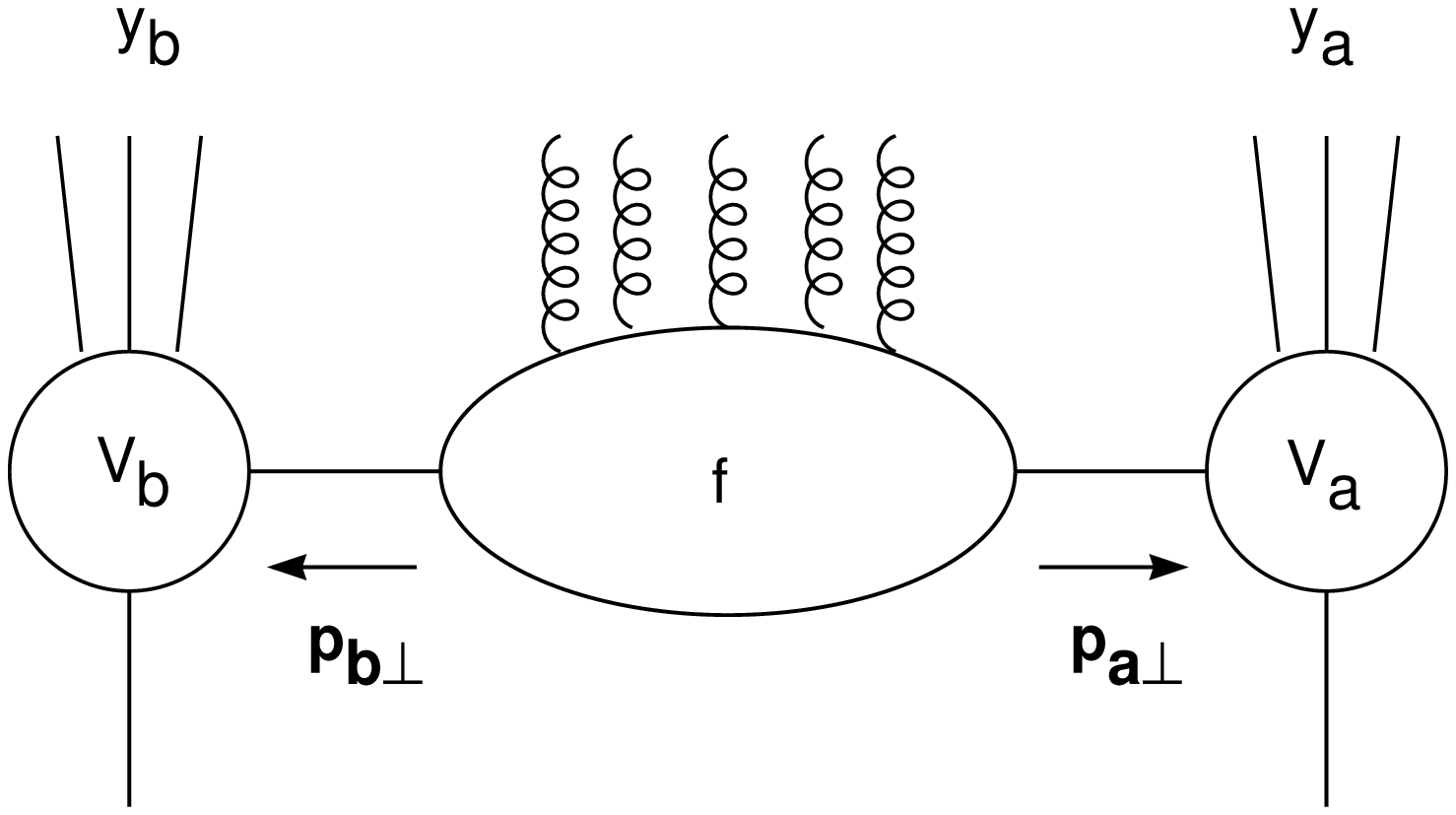}}
\vskip-5.0cm
\vskip6pt
\baselineskip=12pt
Fig.~1: A schematic picture of the cross section for producing particles
at large rapidity separation.
\vskip-1cm
\epsfysize=16cm
\centerline{\epsffile{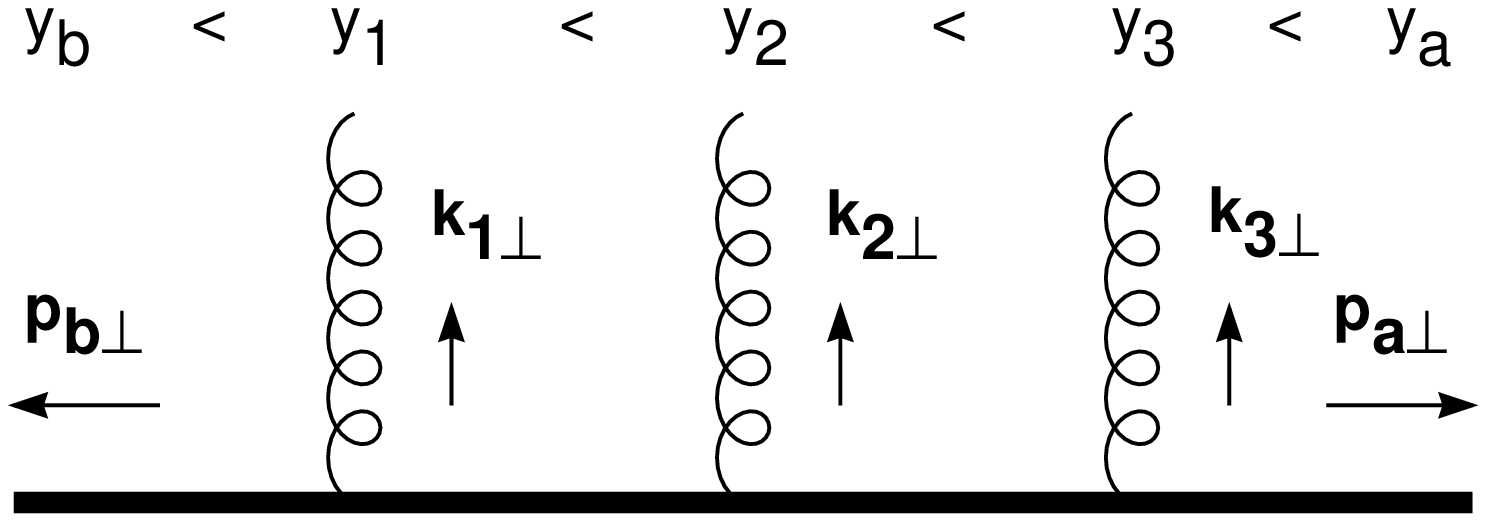}}
\vskip-6cm
\vskip6pt
\baselineskip=12pt
Fig.~2: The amplitude for three gluon emissions in the BFKL
approximation.
\end{figure}

\newpage

\begin{figure}
\vskip-1cm
\epsfysize=8cm
\centerline{\epsffile{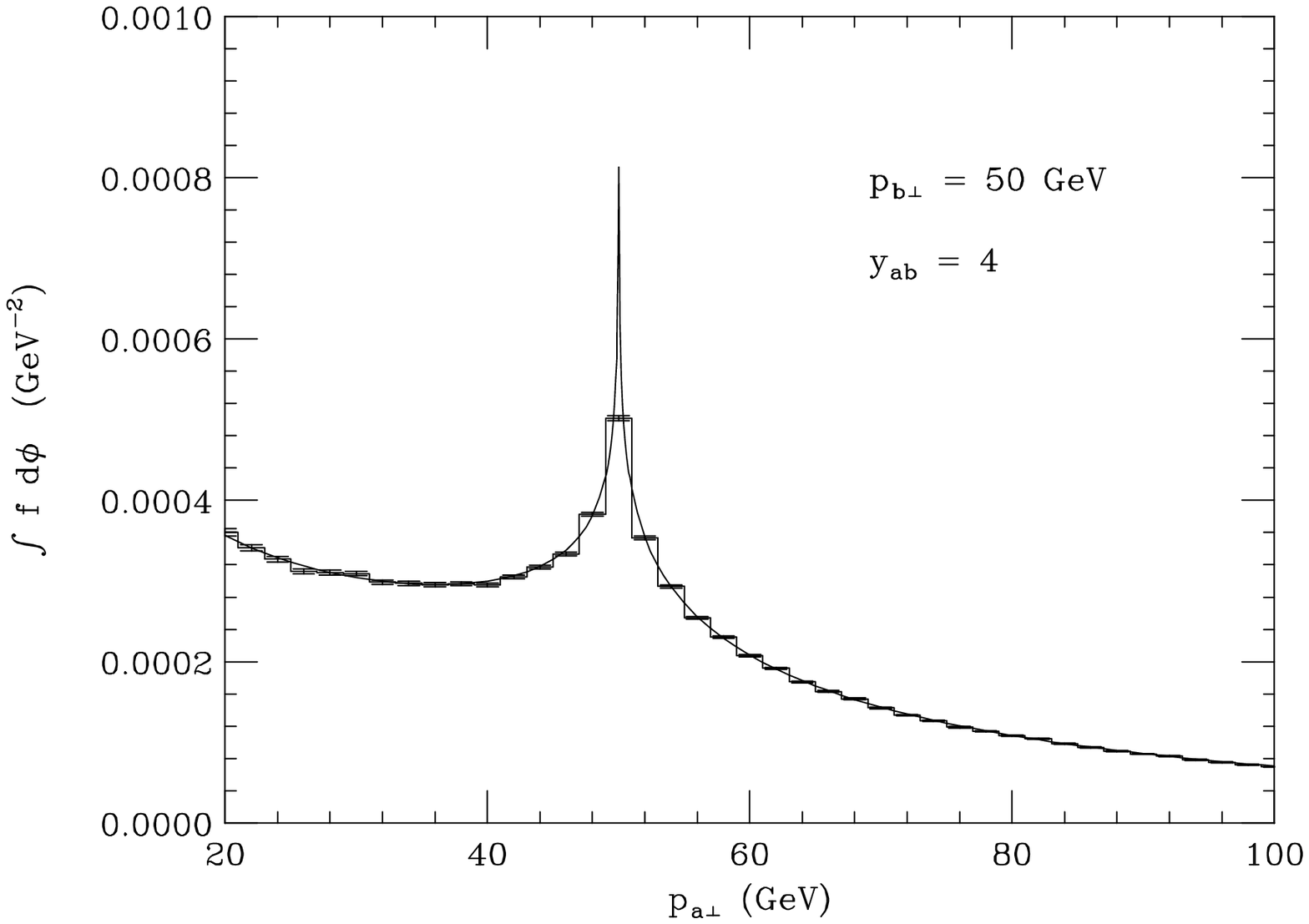}}
\vskip+0cm
\vskip12pt
\baselineskip=12pt
Fig.~3: Comparison of the Monte Carlo solution (histogram) with
the standard solution to BFKL (solid). 
\vskip+1cm
\vskip-6pt
\epsfysize=8cm
\centerline{\epsffile{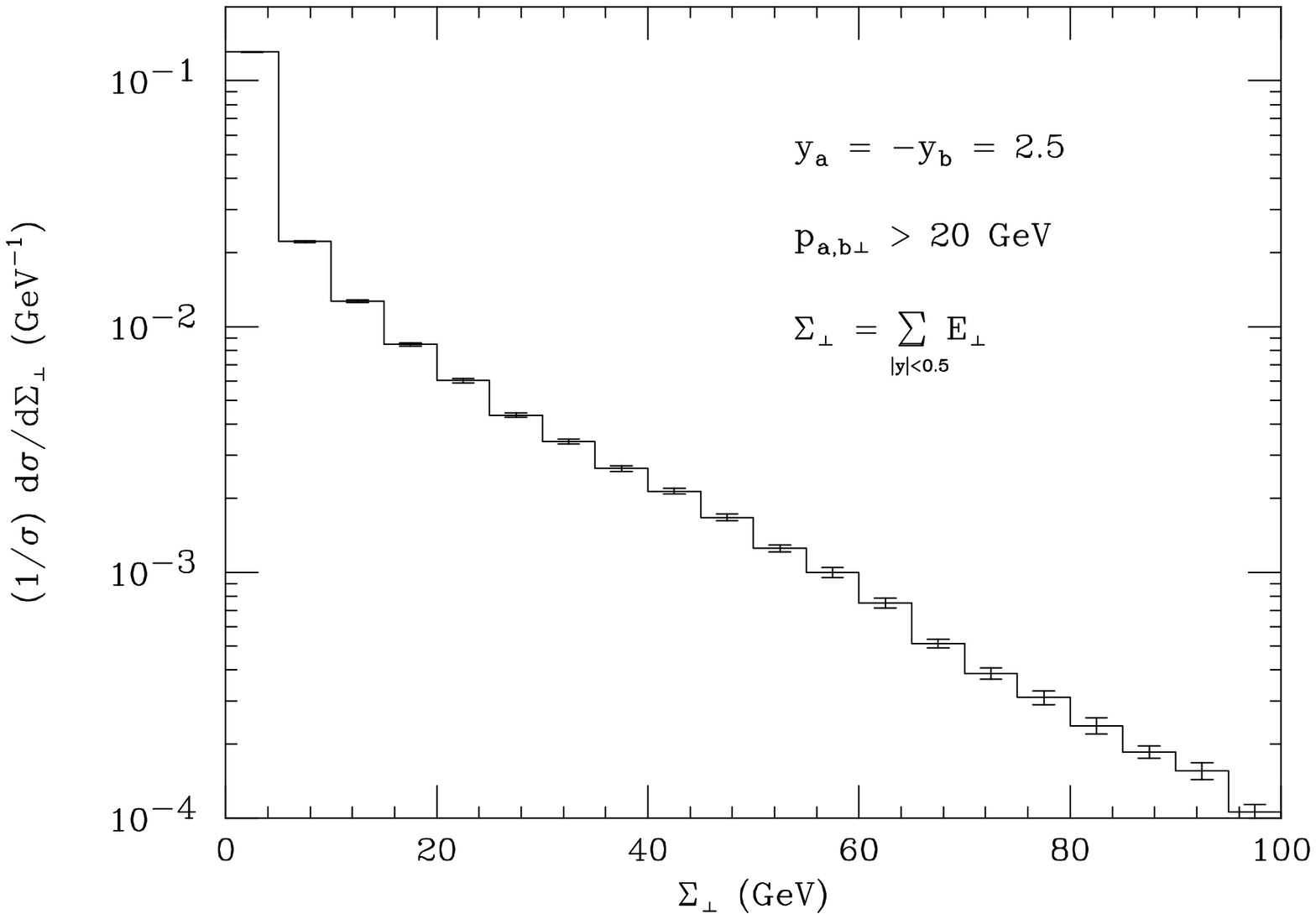}}
\vskip+0cm
\vskip6pt
\baselineskip=12pt
Fig.~4: The distribution of $\Sigma_\perp$ as explained in the
text. 
\end{figure}

\newpage

\end{document}